# Mixed Algorithm of SINDy and HAVOK for Measure-Based Analysis of Power System with Inverter-based Resources

Reza Saeed Kandezy and John Ning Jiang
School of Electrical and Computer Engineering
the University of Oklahoma
Norman, OK, USA
(e-mail: reza.kandezy@ou.edu, jnjiang@ou.edu).

*Abstract*-- **Artificial intelligence and machine learning is enhancing electric grids by offering data analysis tools that can be used to operate the power grid more reliably. However, the complex nonlinear dynamics, particularly when coupled with multi-scale interactions among Inverter-based renewable energy Resources, calls for effective algorithms for power system application.**

**This paper presents affective novel algorithm to detect various nonlinear dynamics, which is built upon: the Sparse Identification of Nonlinear Dynamics method for nonlinear dynamics detection; and Hankel Alternative View of Koopman method for multi-scale decomposition. We show that, by an appropriate integration of the strengths of the two, the mixed algorithm not only can detect the nonlinearity, but also it distinguishes the nonlinearity caused by coupled Inverter-based resources from the more familiar ones caused synchronous generators. This shows that the proposal algorithm can be a promising application of artificial intelligence and machine learning for data measure-based analysis to support operation of power system with integrated renewables.**

*Index Terms*-- **HAVOK, Inverter-based resources, Machine learning, Measure-based method, Model identification, Multi-scale dynamics, Non-linear dynamics, Power system, SINDy.**

## I. Introduction

Although machine learning (ML) algorithm and artificial intelligent (AI) enhanced the analysis of the system by addressing the limitations of traditional models to capture the complex nonlinear dynamics of power systems, the integration of inverter-based energy resources (IBRs) with coupled muti-scale dynamics necessitate an alternative approach utilizing more advanced algorithms to improve modeling accuracy and system optimization [1-4]. In the conventional view, the IBRs technology will introduce nonlinearity into the power system, however, the nonlinearity introduced by the IBRs classified into the module-level nonlinearity thus the power system remains first order, i.e., linearizable [5]. The current state-of-the -art ML algorithm showed remarkable performance regarding model identifications and system analysis of linear and nonlinear first-order system. Increase in the penetration of renewable energy, specifically IBRs, in power system caused a shift in the nature of the system by introducing coupled multi-scale nonlinear dynamic. This changed introduced and interactive complex nonlinearity on system-level to the grid which elevate the power system to the second-order system. The generic ML algorithms struggle to capture systems' multi-scale temporal and spatial complexities, further limiting their accuracy and effectiveness in modeling and analysis of power system derived by IBRs, where fast and slow dynamics capture localized phenomena and overall spatial patterns [5] [6].

Since the introduction of SINDy algorithm in 2016 [7] it has seen widespread use in model identification across various disciplines, showing remarkable performance by explicitly identifying underlying governing equations through sparse regression techniques, which leads to interpretable models while effectively addressing model complexity [7], [8]. The SINDy algorithm has been used in limited number of the field of power system analysis [9-12]. Notably, these studies predominantly analyzed the general power system and relied on first-order system models, as exemplified by [9] in 2020 paper on power system applications.

SINDy is a measure-based method specifically designed to discover governing equations or mathematical models from observed data. Several studies in different disciplines demonstrated the performance and practicality of SINDy in model identifications [13]. By leveraging compressed sensing and sparse modeling principles, SINDy offers robustness and the potential for generalization, allowing for identifying key dynamical features with relatively few measurements. However, it's important to note that SINDy can be sensitive to noise. At its core, the generic SINDy assumes the system as linearizable system, i.e., first-order nonlinear system. Therefore, generic SINDy faces challenge in coupled multi-scale systems where multiple variables interact strongly [7], [8]. These systems often involve intricate relationships, and accurately capturing these dynamics requires disentanglement of the multi-scale dynamics [14]. This intricate and multifaceted nonlinear effects of integrating IBRs and renewables, manifesting as second-order nonlinearity, was not considered in [9-12].

HAVOK is a decomposition technique approximating chaotic dynamics as an intermittently forced linear system, combining principles from delay embedding and Koopman theory [14]. It leverages real-time data covariance analysis to uncover the underlying structures within high-dimensional data. The method employs time-delay embeddings to capture latent variables and intrinsic measurement coordinates, rooted in the theory of Koopman operators [14]. By describing the evolution of observables as an infinite-dimensional linear process, Koopman operators offer an alternative perspective for analyzing dynamical systems without explicitly solving their underlying differential equations. The HAVOK method extends this framework by utilizing time-delay embeddings and intrinsic measurement coordinates, enabling the disentanglement of intricate couplings in multiscale systems.

The combination of SINDy with HAVOK addresses the challenges posed by complex coupled multi-scale interactions and dynamics within system such as power grids integrated with intermittent IBRs. This integrated approach leverages SINDy to identify the underlying governing equations and utilizes HAVOK to exploit the Hankel matrix structure of the data, enabling the extraction of informative patterns and features that characterize the power grid's behavior.

The contributions of our work are summarized in two aspects,
1. We present a novel mixed algorithm that can be used for data-based analysis. This algorithm combines the advantages of two powerful methods recently developed in the field of data science for the detection and analysis of complex multi-scale nonlinear dynamics in power systems with IBRs.
2. Through a demonstrative study showing the effectiveness of identifying various nonlinear dynamics with different characteristics, we demonstrate, in a broader sense, the necessity of integrating different AI/ML data analytics to the development of more effective measurement-based tools for power system analysis.

The subsequent sections of this manuscript are organized as follows: Section II describe the developed framework of generic SINDy method and supporting HAVOK decomposition in power system. Section III demonstrates the proposed methods' performance by evaluating and examining the obtained outcomes. The final segment comprises the concluding remarks, emphasizing the results' significance and potential research directions for future studies.

## II. SINDy Algorithm and Supporting HAVOK Decomposition

### A. Developed method based on generic SINDy algorithm:

The dynamics of power system can be described by the following general form [15]:

$$\frac{dv(t)}{dt} = f(v(t)), \quad (1)$$

where $v(t) \in R^n$ represents the system's voltage at time $t$ and $f(v(t))$ encompasses the dynamic constraints governing the system's equations, including parameters, time dependence, and external forcing. To determine the function $f$ from available data, a time history of the system's voltage, denoted as $v(t)$, is collected. The derivative of $v(t)$, denoted as $\dot{v}(t)$, is directly or numerically approximated. The data is sampled at various time instances, $\{t_1, t_2, \ldots, t_m\}$ and organized into $V$ and $\dot{V}$ matrices:

$$V = \begin{bmatrix} | & | & & | \\ v(t_1), v(t_2), \ldots, v(t_m) \\ | & | & & | \end{bmatrix}^T, \quad (2)$$

$$\dot{V} = \begin{bmatrix} | & | & & | \\ \dot{v}(t_1), \dot{v}(t_2), \ldots, \dot{v}(t_m) \\ | & | & & | \end{bmatrix}^T. \quad (3)$$

The next step in the approach involves defining a library of candidate nonlinear functions, denoted as $\Theta(V)$, where $V$ is the data matrix that contains observed data points of the system voltage. These functions can include polynomials, trigonometric, exponentials, logarithmic functions, and other suitable nonlinear expressions.

$$\Theta(V) = [1, V, V^2, V^3, \ldots, sin(V), cos(V), \ldots]. \quad (4)$$

Assuming that only a few of these nonlinearities are active in each row of $f$, a sparse regression problem is formulated to determine the sparse vectors of coefficients,

$$\Xi = \begin{bmatrix} | & | & & | \\ \xi_1, \xi_2, \ldots, \xi_n \\ | & | & & | \end{bmatrix}, \quad (5)$$

which indicate the active nonlinearities. Mathematically, this can be expressed as:

$$\dot{V} = \Theta(V)\Xi. \quad (6)$$

Given the voltage matrix $V$ and the library of candidate functions $\Theta(V)$, The method formulates the sparse regression problem as follows:

$$minimize \; ||\dot{V} - \Theta(V)\Xi||_2 + \lambda \, ||\Xi||_1 \quad (7)$$

where $\Xi$ is the sparse coefficients representing the importance or relevance of each term in the library, $|| \cdot ||_2$ denotes the L2 norm, $|| \cdot ||_1$ represents the L1 norm, and $\lambda$ is a regularization parameter that controls the trade-off between data fidelity and sparsity. The first term ensures that the model predictions, obtained by multiplying $\Theta(V)$ with $\Xi$, are close to the observed data, while the second term encourages a sparse solution by promoting a minimal number of nonzero coefficients [7].

### B. Adaptation of HAVOK decomposition on power system:

Consider a record of the voltage on a single point $v(t)$, where we generate a delay embedding vector denoted as $V(t)$. This vector is comprised of delayed measurements of $v(t)$, at different time points and can be formally expressed as:

$$V(t) = [v(t), v(t - \tau), v(t - 2\tau), \ldots, v(t - (m-1)\tau)]^T, \quad (8)$$

where $\tau$ signifies the time delay between measurements and m corresponds to the embedding dimension.

We conduct the Singular Value Decomposition (SVD) of the matrix $V$, which is constructed from the time series of a single measurement voltage $v(t)$ to obtain intrinsic measurement coordinates. The SVD of V yields the following decomposition:

$$V = Y\Sigma U^*, \qquad (9)$$

$Y$ and $U$ represent orthogonal matrices, the $\cdot^*$ represent conjugate transpose, and $\Sigma$ is a diagonal matrix comprising singular values. $Y$ columns correspond to the eigen-time-delay coordinates, which capture the essential structure of the system's dynamics. The resultant linear model within the HAVOK method can be succinctly expressed as:

$$\frac{d}{dt} u(t) = Au(t) + Bu_r(t), \qquad (10)$$

where $u(t)$ signifies a vector comprising the first $r-1$ eigen-time-delay coordinates, $A$ denotes a matrix capturing the linear dynamics, and $B$ represents a matrix characterizing the coupling between the eigen-time-delay coordinates and the forcing term $u_r(t)$. In the case of chaotic systems, the linear model based on the first $r-1$ terms provide a commendable approximation, while the forcing term encapsulates the nonlinear and chaotic behavior of coupled multi-scale dynamic. It has to be noted that in scenarios where the intervals between multi-scale dynamics' time constants are significantly disparate, the necessity for sampling becomes considerably demanding [16]. By employing burst sampling, the data requirements for SINDy remain relatively consistent, even as the temporal scales diverge [17]. This approach reduces the sampling rate, which proves advantageous when limitations exist on the number of samples that can be acquired due to bandwidth restrictions [16].

III. DEMONSTRATION STUDY

In this paper, conducted on an IEEE 15-bus power grid (depicted in Figure 1), the developed method, based on generic SINDy algorithm and HAVOK decomposition, was deployed to scrutinize voltage waveforms and uncover system dynamics. The algorithm's performance was assessed across diverse and intricate operational scenarios. The deliberate selection of this particular power system configuration serves as a foundation for showcasing SINDy's performance under varying conditions and highlights the practicality of HAVOK in tackling the challenges presented by complex, multi-scale dynamics. The evaluation of the method in the large-scale systems and further experimental analysis are reserved for future investigations.

A. Test setup:

The system architecture consists of 15 buses, representing distinct nodes within the power system network, and these buses are interconnected through branches that symbolize the power transmission lines. Each bus has a unique set of parameters and attributes and is connected to neighboring buss via branches characterized by specific impedance, which govern the dynamics of power flow among the interconnected buses. Table 1 provides a comprehensive overview of the network configuration and its branch parameters.

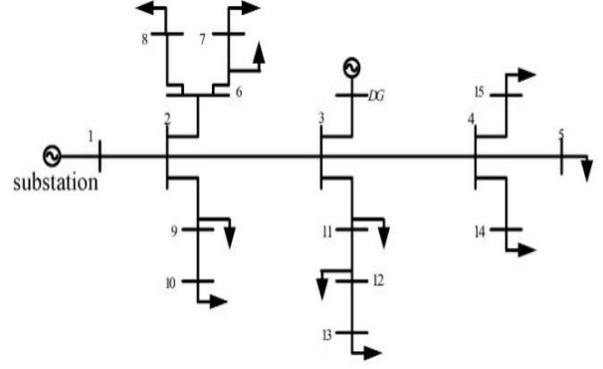

Figure 1- Single line diagram of implemented IEEE 15 bus network

Table1 – Implemented IEEE 15 bus network configuration

| Line index | From bus | To bus | $r+xi$ (Ω) | Node index | $P_{load}+Q_{load}$ (kW+kVAR) |
|---|---|---|---|---|---|
| 1 | 1 | 2 | 1.53+1.778i | 2 | 100+60 |
| 2 | 2 | 3 | 1.037+1.071i | 3 | 90+40 |
| 3 | 3 | 4 | 1.224+1.428i | 4 | 120+80 |
| 4 | 4 | 5 | 1.262+1.499i | 5 | 60+30 |
| 5 | 5 | 9 | 1.176+1.335i | 6 | 60+20 |
| 6 | 6 | 10 | 1.1+0.6188i | 7 | 200+100 |
| 7 | 7 | 6 | 1.174+0.2351i | 8 | 200+100 |
| 8 | 8 | 7 | 1.174+0.74i | 9 | 60+20 |
| 9 | 9 | 8 | 1.174+ 0.74i | 10 | 60+20 |
| 10 | 10 | 11 | 1.15+ 0.065i | 11 | 45+30 |
| 11 | 11 | 12 | 1.274+1.522i | 12 | 60+35 |
| 12 | 12 | 13 | 1.274+1.522i | 13 | 60+35 |
| 13 | 13 | 14 | 1.075+1.522i | 14 | 120+80 |
| 14 | 14 | 15 | 1.075+ 1.522i | 15 | 60+10 |

This investigation explores a comprehensive set of power system conditions, encompassing both abrupt changes (faults) and gradual changes (load variations), in the context of conventional synchronous generators (SG) and IBRs. The study encompasses three distinct scenarios, representing both single- and multi-dynamic systems. The first scenario examines a system solely supplied by a synchronous generator at Bus 1 and Bus 3. The second scenario incorporates the integration of an IBR at Bus 3, sharing the load demand equally with the synchronous generator at Bus 1. In the third scenario, the network is subjected to a 100% penetration of IBRs, where the demand is supplied by two IBRs located at Bus 1 and Bus 3. Each scenario spans 10 seconds, with the synchronous generators and IBRs initiated at $t = 0s$. At $t = 3.3s$, a three-phase-to-ground fault occurs at Bus 10, cleared after four cycles of the fundamental frequency (60 Hz). Furthermore, at $t = 7s$, an extra load is connected to Bus 14, disconnected at $t = 8s$.

The simulation duration of 10 s and total sample count of 200,000 (20,000 sample per second) are determined to capture temporal dynamics faithfully. Furthermore, a polynomial library is constructed with a precise third order polynomial and regularization parameter (0.8).

In multiscale dynamic conditions, the SINDy enhanced by HAVOK decomposition were applied. The essential parameters such as the number of samples per burst (1000 samples per burst), the optimal hard threshold value for singular values ($\frac{4}{\sqrt{3}}$) [18], and the sampling time interval (50 microseconds) play crucial roles in this process.

## B. Model identification using generic SINDY

### 1) General power system (SG driven):

The results in the first scenario demonstrate SINDy's accuracy in estimating the system model. For the steady-state dynamics of the power system supplied by SG, the sparse model adeptly replicates the dynamics observed in measurements. As depicted in Figure 2.a, the reconstructed waveform resulting from identified model closely aligns with the actual data collected from the power network, all presented per unit. Notably, the algorithm not only correctly identifies the terms governing the dynamics but also accurately determines the associated coefficients, with deviations well within a remarkable 0.03%.

Continuing in this scenario, involving sudden and gradual changes to system parameters in transient states, the generic algorithm captures the system nonlinear dynamics and tracks changes, while experiencing a minor fluctuation during these transitions, resulting in slight increases in approximation errors. However, the results shows that the generic algorithm swiftly recovers its accuracy once changes are detected, as evident in faults and load variations, shown in Figure 2.b and Figure 2.c, respectively.

### 2) Power system integrated with IBRs:

When IBRs were introduced into the power grid, the performance of generic algorithm in model identification deteriorated since it could not distinguish the multi-scale nonlinear dynamics. The result obtain in this scenario indicates that the algorithm's errors experience a notable increase.

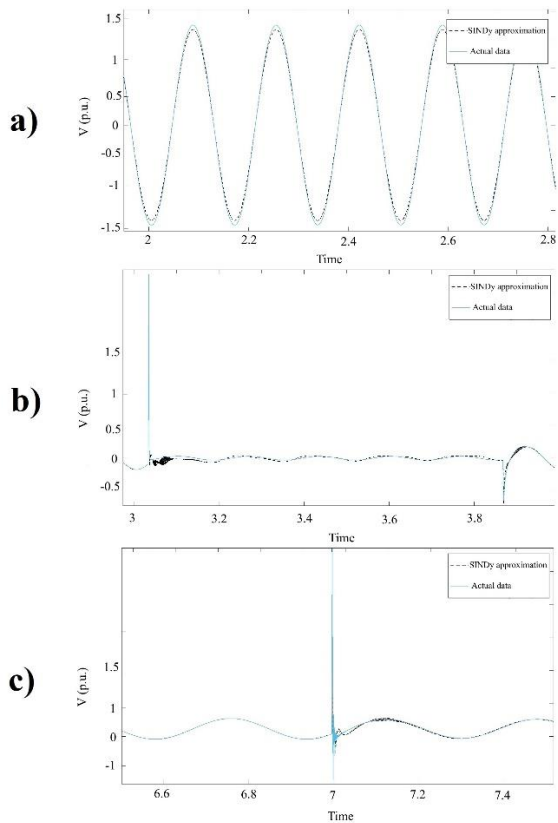

Figure 2- The actual measurement and generic algorithm approximation during a) steady state, b) fault and c) load change in power system supplied by SG

Looking into the detail, by comparing the voltage variations approximated through the generic algorithm and the actual measurement it is observed that the generic algorithm failed to distinguish the nonlinearity caused by fast dynamics, in coupled multi-scaled dynamics. As presented in Figure 3, the generic algorithm did not recognize the fast dynamics in the voltage. Detailed analysis of the approximated voltage from, shown with dash lines in Figure 3, indicates the inclusion of harmonic distortions in the final result, comparing to the voltage waveform approximated in previous scenario, i.e., first-order system. Furthermore, the phase, frequency and amplitudes of the approximated voltages shows deviation from cycle to cycle.

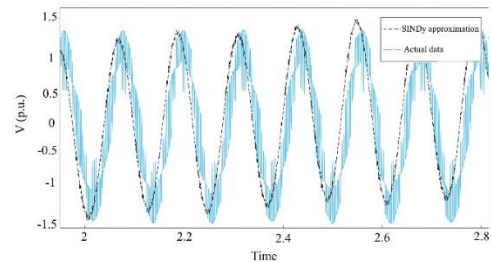

Figure 3- The actual dynamic and generic algorithm approximation during steady state in power system supplied by SG and IBRs.

The investigation continued by increase of the IBR penetration to 100 % and the result, obtained from generic algorithm was subjected to the same analysis. After analyzing the voltage waveform approximated by generic algorithm in second and third scenario the error from actual data was calculated. The findings indicate 50 % penetration of IBRs in power system led to a five-fold increase in errors compared to scenarios with conventional SG, while achieving a 100% penetration of IBRs resulted in approximately seven times higher errors, as illustrated in Figure 4.

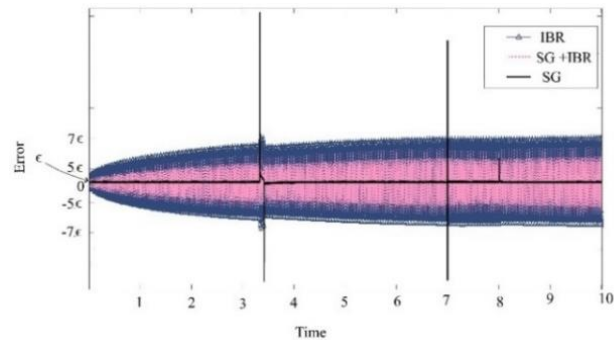

Figure 4- The errors between actual dynamic and generic algorithm approximation during steady state and transient in power system

The introduction of harmonic distortion and deviation in voltage, frequency and phase by the generic algorithm in power its approximation is due to inability of recognizing the fast dynamics. The generic algorithm could not identify the second-order nonlinearity in the system, hence, translate it into inaccurate level of first-order nonlinearity showing that the presence of IBRs in the power grid introduces complexities or dynamics not adequately captured by generic algorithm.

*C. Model identification using enhanced mixed algorithm (generic SINDy and HAVOK decomposition):*

The second and third scenarios were repeated using the enhanced algorithm, using combination of generic SINDy and HAVOK decomposition. The voltage was approximated using the mixed algorithm and has been compared to the actual data in both scenarios. Figure 5 present the approximated voltage and the actual measurement for the voltage in the third scenario (100% penetration of IBRs) for steady and transient states. The errors calculated for the second and third scenarios were 1.14 and 1.79 times of the base error ($1.14\epsilon$ and $1.79\epsilon$ where $\epsilon$ is the error captured in the first scenario).

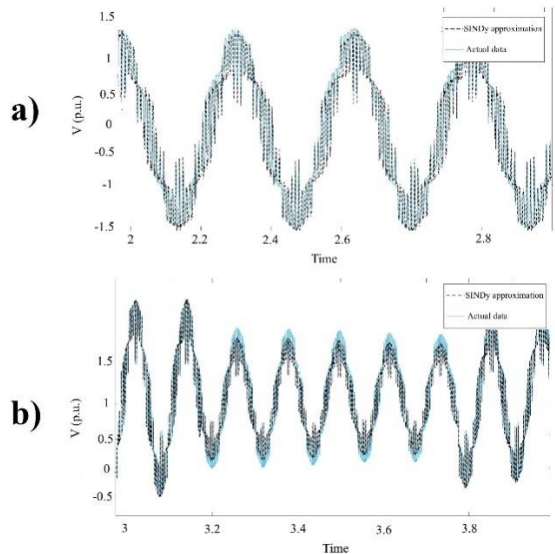

Figure 5- The actual measurement and mixed algorithm approximation during a) steady state and b) transient state in power system supplied by IBR

The results demonstrate a notable enhancement in the approximation of the power system with second-order nonlinearity. The mixed algorithm was able to recognize the fast dynamic within the coupled multi-dynamic systems. The enhanced algorithm was able to capture and follow the dynamic in steady state, and transient state, shown in Figure 5.a and Figure 5.b respectively. Note that the absence of voltage fluctuations in the fault incident are due to the limitation of the IBRs in providing fault current, leading to fast and hardly recognizable impulses. This indicates that the combined method can characterize the complex interactions and dynamics present in the system.

## IV. CONCLUSION

This paper presents the concept and generic framework of a mixed algorithm that could be used for measurement-based power system analysis. To address the challenging impact of coupled multi-scale dynamics of IBRs on the complex nonlinear dynamics experienced in the power grid, we integrate the HAVOK decomposition method with the powerful SINDy analytics recently developed in data science, along with a set of illustrative case studies in power system analysis.

The case studies not only demonstrate the effectiveness of detecting various nonlinear dynamics in power systems but also clearly demonstrate the promising capability of the mixed algorithm to separate the nonlinear dynamics induced by coupled IBRs from those caused by synchronous machines.

Although the effectiveness of the algorithm is demonstrated with case studies using a small-scale IEEE model, the benefits of integrating various data analytic tools are clearly shown from a broader perspective.

Currently, we are collaborating with power and utility companies to develop larger-scale models and obtain real data from measurements to apply the mixed algorithm in building measurement-based tools for detecting nonlinear dynamics, such as sub-synchronous oscillations and resonance, as well as complex transient energy waves caused by inverter interactions.

## V. ACKNOWLEDGMENT

We express our profound appreciation to all those who contributed directly or indirectly to the successful completion of this research paper with their valuable time, insights, recommendations, and support.